\documentclass[twocolumn,aps,superscriptaddress]{revtex4}
\usepackage{epsfig}
\usepackage{amsmath,latexsym,amssymb,amsfonts}
\newcommand{\be}{\begin{equation}}
\newcommand{\ee}{\end{equation}}
\newcommand{\bea}{\begin{eqnarray}}
\newcommand{\eea}{\end{eqnarray}}

\newcommand{\bc}{\begin{center}}
\newcommand{\ec}{\end{center}}

\renewcommand{\(}{\left(}
\renewcommand{\)}{\right)}
\renewcommand{\[}{\left[}
\renewcommand{\]}{\right]}
\newcommand{\forget}[1]{}
\newcommand{\half}{\frac{1}{2}}

\begin{document}

\preprint{}
\title{Detecting separable states via semidefinite programs}
\author{Federico M. Spedalieri}
\email{federico@ee.ucla.edu}
\affiliation{Department of Electrical Engineering, University
of California, Los Angeles, Los Angeles, California 90095}

\date{\today}

\begin{abstract}
We introduce a new technique to detect separable states using
semidefinite programs. This approach provides a sufficient
condition for separability of a state that is based on the
existence of a certain local linear map applied to a known separable
state. When a state is shown to be separable, a proof of this
fact is provided in the form of an explicit convex decomposition of the
state in terms of product states. All states in the interior of the set
of separable states can be detected in this way, except maybe for a set
of measure zero. Even though this technique is
more suited for a numerical approach, a new analytical criterion
for separability can also be derived. 

\end{abstract}

\maketitle

\section{Introduction}

Entanglement is one of the most important resources for
quantum information processing. It allows us to perform 
such tasks as teleportation, secure quantum key distribution,
superdense coding and quantum computation to name a few~\cite{nielsen2000}.
Because of its central role a great deal of effort has been
put into its characterization in the past few years.

One of the most important questions, and also one very difficult
to answer in general, is whether a given mixed state is entangled
or separable. The decision problem associated with this characterization
has been shown to be NP-hard~\cite{gurvits2003a}, so a simple
practical procedure to answer this question is not likely to exist.
To tackle the problem, many incomplete criteria have been developed
in the form of either necessary or sufficient conditions for separability.
Several of these criteria are based on relatively easy to verify
properties that separable states must satisfy. If a given state
fails such a test, it must be entangled. However, if the state passes the
test, the result is inconclusive. This type of criteria includes
the positive partial transpose (PPT) 
criterion~\cite{peres1996a,horodecki1996a} and its 
extensions~\cite{doherty2002a,doherty2003d},
the reduction criterion~\cite{cerf1999a} and the range 
criterion~\cite{horodecki1997a} among others.

In this paper we introduce a new criterion for separability that
works in a complementary way. If a given state passes a certain
test, the state must be separable. Furthermore, a proof of this
fact is given as an explicit convex decomposition of the state in
terms of product states. The criterion is based on the fact that
separable states preserve their separability property 
when a local map with certain properties acts on them. This
local map need not be a physically implementable map, or even
a positive map. The key point is that the search for such
map can be implemented as a semidefinite program (SDP),
which is a class of convex optimization problems for which 
efficient algorithms are known~\cite{VaB:96}. Semidefinite programs have found
widespread application in quantum information~\cite{rains2001a,doherty2003d,
brandao2004a,fletcher2006a}. 
Even though this new criterion is formulated in a way that is
best suited for a numerical approach, it can also be used to
derive a new analytical condition for separability. 

The paper is organized as follows. In Section I we present the
basic idea of our approach by looking at the action of local 
maps on separable states. In Section II we introduce the concept 
of base states and discuss the properties they must have. Section III
introduces an improved approach that can detect more separable states.
Section IV deals with the problem of characterizing the set of states that
are detected by this procedure. Section V shows how to extract
a new analytical criterion for separability and serves as an illustration
of the technique. Finally, the conclusions are presented in Section VI.

\section{Local maps on separable states}

Let $\rho_0$ be a separable state on ${\cal H}_A \otimes {\cal H}_B$, where
for now we will consider that $\mathrm{dim}\, {\cal H}_A = 
\mathrm{dim}\, {\cal H}_B = d$. Then, by definition, we 
know that $\rho_0$ can be written as a convex combination of product states, 
namely
\be
\label{convdec}
\rho_0 = \sum_i p_i \,\rho_A^{(i)} \otimes \rho_B^{(i)},
\ee
with $0 \le p_i \le 1, \sum_i p_i =1$ and $\rho_A^{(i)}$ and $\rho_B^{(i)}$
are density matrices in ${\cal H}_A$ and ${\cal H}_B$ respectively. 
Suppose that we have a linear map
$\Lambda$ on density matrices over ${\cal H}_B$, that satisfies
$\Lambda (\rho_B^{(i)}) \ge 0$ for all $i$ (where the inequality means
the matrix is positive semidefinite (PSD)). Then, if we apply the map
${\cal I} \otimes \Lambda$ to $\rho_0$, the resulting state $\rho$ is,
up to some normalization constant,
also a separable state, since
\be
\label{basicobs}
\rho = {\cal I} \otimes \Lambda (\rho_0) = \sum_i p_i \,\rho_A^{(i)} \otimes 
\Lambda(\rho_B^{(i)}),
\ee
gives a decomposition of $\rho$ as a convex combination of product states.
Note that $\Lambda$ need not be a positive map, since it is only required
to be positive over a (typically) finite set of density matrices
over ${\cal H}_B$. Furthermore, since $\Lambda$ may not even be normalized
(in the sense that $\mathrm{Tr}[\Lambda(\rho_B^{(i)})]$ may not be equal
to 1), after the appropriate normalization the probabilities $p_i$ in the decomposition of $\rho$ may be different from those in the
decomposition of $\rho_0$. 

Our approach to identifying separable states will be based precisely
on this simple observation. Basically, given a state $\sigma$, if we find a map
$\Lambda$ satisfying the above mentioned criterion, and such that
$\sigma = {\cal I} \otimes \Lambda (\rho_0)$, we can assure that 
$\sigma$ is a separable state. Furthermore, we will be able to provide
a proof of this fact in the form of an explicit decomposition of
$\sigma$ as a convex combination of product states. We will see that
the search for this map corresponds to solving a linear system and
checking for the appropriate positivity requirements afterwards. 
Then we will show that we can slightly modify our basic observation
(\ref{basicobs}) to enlarge the set of maps that would allow us to detect
separable states, and we will show that this new search can be cast as
a semidefinite program. 

It is clear that for this approach to be useful, we need to start
with a separable state $\rho_0$ for which we know an explicit decomposition
of the form (\ref{convdec}). We will refer to such a state as a 
\emph{base state}, since it is used as a ``base'' from which to reach
other separable states. The usefulness of this technique
depends on how the state $\rho_0$ is chosen, and we will start by 
discussing what 
properties it must have. 
     
\section{Base states}

Given a state $\sigma$ the question of whether it satisfies
$\sigma = {\cal I} \otimes \Lambda (\rho_0)$ for some map $\Lambda$ can
be answered by solving a linear system of equations. Both $\sigma$ and 
$\rho_0$ can be decomposed with respect to a basis for density matrices
over ${\cal H}_A \otimes {\cal H}_B$ as
\be
\rho_0 = \sum_{m,n,r,s = 1}^d R_{mnrs} \, |m\rangle \langle n| \otimes
|r\rangle \langle s|,
\ee
and
\be 
\sigma = \sum_{m,n,r,s = 1}^d S_{mnrs} \, |m\rangle \langle n| \otimes
|r\rangle \langle s|.
\ee
We can also define a basis on the linear space of maps over operators in
${\cal H}_B$ given by the set of maps $\{ \Lambda_{ijkl} \}_{i,j,k,l=1}^d$ 
that satisfy
\be 
\Lambda_{ijkl} (|r\rangle \langle s|) = |k\rangle \langle l|\, \delta_{ir}
\,\delta_{js}.
\ee
The most general linear map can then be written as $\Lambda = \sum_{i,j,k,l=1}^d
 x_{ijkl} \,\Lambda_{ijkl}$, for some arbitrary coefficients $x_{ijkl}$. 
Then the equation 
$\sigma = {\cal I} \otimes \Lambda (\rho_0)$
reduces to a system of linear equations given by
\be
\label{linsys}
S_{mnkl} = \sum_{r,s=1}^d R_{mnrs} \,x_{klrs}.
\ee
This is a system of $d^4$ equations with $d^4$ unknowns. It will
always have a solution, but we also need to impose on this solution the
positivity conditions $\Lambda (\rho_B^{(i)}) \ge 0, \forall i$. If the
linear system (\ref{linsys}) has a unique solution, we can check these
conditions by computing eigenvalues. If the system is undetermined
we can solve for some of the coefficients $x_{ijkl}$ and write
$\Lambda = \Lambda_0 + \sum_J \Lambda_J$, where $\Lambda_0$ is the fixed
part of the solution and the sum groups the remaining terms. Thus the problem
reduces to checking whether the map $\Lambda$ is positive over
the set $\{ \rho_B^{(i)} \}$. This can be written as a linear matrix 
inequality (LMI), and so we can cast our problem in the form of the 
semidefinite program (SDP)
\bea
\label{SDP}
\mathrm{min}&  0 \nonumber\\
\mathrm{subject\ to}& \bigoplus_i \Lambda_0 (\rho_B^{(i)}) + \sum_J x_J
 \bigoplus_i
 \Lambda_J (\rho_B^{(i)}) \ge 0.
\eea
The size of the SDP will be determined by the dimension $d$ and the number
of terms in the decomposition (\ref{convdec}) of $\rho_0$. Caratheodory's
theorem~\cite{rockafellar1970a} assures that any separable 
state can be decomposed as a 
convex combination of at most $d^4 +1$ product states, so the size
of the SDP remains polynomial in $d$. We will see later that we
would have ample freedom to choose the base state $\rho_0$, and so
we can keep the size of the SDP under control.

The uniqueness of the solution of the system (\ref{linsys}) depends
an whether the matrix of the system is invertible or not. From (\ref{linsys})
we can see that this matrix will depend only on the components $R_{mnrs}$, and
hence it is completely determined by the base state $\rho_0$. From this
point of view it is a bit cumbersome to determine whether a given base
state makes the system (\ref{linsys}) invertible or not. Fortunately there is
an alternative way of looking at this problem that connects with previous
work and provides us with a nicer and more practical 
characterization of base states.

First, let us note that for a fixed base state $\rho_0$ we can regard 
the expression ${\cal I} \otimes \Lambda (\rho_0)$ as a function that assigns
to every map $\Lambda$ the operator ${\cal I} \otimes \Lambda (\rho_0)$.
Note that we refer to it as an operator and not a state, since we have
imposed no conditions on $\Lambda$, which may not be positive. If we 
choose $\rho_0 = |\psi^+ \rangle \langle \psi^+|$ (with $|\psi^+\rangle
=\frac{1}{\sqrt{d}} \sum_{i=1}^d |ii\rangle$ the maximally entangled
state), then we know that the linear function $f(\Lambda) = 
{\cal I} \otimes \Lambda (|\psi^+ \rangle \langle \psi^+|)$ has very important
and useful properties. It defines an isomorphism between the set of
completely positive (CP) maps, mapping states over ${\cal H}_B$ to states
over ${\cal H}_A$, and the set of
states over ${\cal H}_A \otimes {\cal H}_B$. This is known as the
\emph{Jamio{\l}kowski isomorphism}~\cite{jamiolkowski1972a}. 
In \cite{d'ariano2003a} D'Ariano and Lo Presti
studied whether other bipartite states besides the maximally entangled
state will induce a one to one mapping between CP maps and states. 
They found that
there are indeed such states and derived a very simple test to identify
them: a state $\rho_0$ will induce such a mapping if the operator
$\check{\rho}_0 = (E \rho_0)^{T_B} E,$ is invertible, with $E=\sum_{ij} 
|ij\rangle\langle ji|$ the swap operator. They called such a state
a \emph{faithful} state. Furthermore, from this characterization
it is easy to see that the set
of faithful states is \emph{dense} over the set of all states, and so
we can have faithful states that are separable. However, we should note
that separable pure states cannot be faithful. In fact, if
$\rho_0 = |x\rangle \langle x| \otimes |y\rangle \langle y|$
(where $|x\rangle = \sum_i x_i |i\rangle$ and $|y\rangle = \sum_i
y_i |i\rangle$), then
$\check{\rho}_0 = |y\rangle \langle x^\ast| \otimes |y^\ast \rangle
\langle x|$, which is clearly not invertible. 

If we use a faithful state as a base state, the function $f$ is a
one to one mapping between CP maps and positive semidefinite operators.
In terms of the linear system (\ref{linsys}) this means that
the system must be invertible since the relationship between states and
maps must be one to one. Similarly, if the system (\ref{linsys}) is 
invertible the state $\rho_0$ must be faithful. It is clear then that
if we use a faithful state as a base state, a larger set of separable
states can be detected than if we use nonfaithful states.

\section{Geometric picture and improved algorithm}
\label{enhSDP}

There is a nice geometric picture of our technique to detect
separable states. First let us note that the set
of all states is the intersection of the cone of positive
semidefinite matrices (known as the PSD cone) and the hyperplane 
defined by the normalization condition $\mathrm{Tr}[\rho] =1$. 
The normalization does not affect the separability property
of a state, so in order to make our discussion simpler we will
sometimes analyze and prove results in terms of the cone structure 
of a set of states. It will be straightforward to understand those
results in terms of the actual set of states satisfying the
normalization condition $\mathrm{Tr}[\rho] =1$.

Let us assume that
we have a faithful base state $\rho_0$ that is separable and has 
a convex decomposition given by (\ref{convdec}). The set of maps 
$K_B = \{\Lambda : \Lambda (\rho_B^{(i)}) \ge 0, \forall i\}$, is a \emph{cone}
(that is, a set closed under linear combinations with nonnegative 
coefficients). The linear map $g$ given by $g(\Lambda) = {\cal I} \otimes 
\Lambda (\rho_0)$, maps $K_B$ into another cone 
contained in the set of separable states (which is itself a cone). 
It is not difficult to see that
this image cone has nonzero measure. First note that $K_B$ 
contains the cone of CP
maps which is in one to one correspondence with the PSD cone 
in ${\cal H}_A \otimes {\cal H}_B$ (the cone 
associated with bipartite density matrices) 
via the linear map $f$ defined in the previous section. 
Both the cone of CP maps and 
the PSD cone are embedded in vector spaces of the same dimension $d^4$,
and since the PSD cone has nonzero measure, it follows that so do
the CP cone and $K_B$.
Since the map $g$ is linear and one to one, it will send a nonzero measure
set into a nonzero measure set.

We have seen that our approach reduces to solving a linear system
and checking certain positivity properties of the solution. In particular
if we use a faithful state as a base state, such linear system
has a unique solution, so no semidefinite program is required.
However, we can improve our technique by considering a 
slightly more general type
of mapping. Instead of considering $g(\Lambda) = {\cal I} \otimes 
\Lambda (\rho_0)$ we can analyze the more general linear map
\be
\label{extendmap} 
\sigma =    
\[{\cal I} \otimes \Lambda_B + \Lambda_A \otimes {\cal I}\] (\rho_0),
\ee
with $\Lambda_B \in K_B$ and $\Lambda_A \in K_A$, 
where $K_A = \{\Lambda_A : \Lambda_A (\rho_A^{(i)}) \ge 0, \forall i\}$.
The maps of the form ${\cal I} \otimes \Lambda_B + \Lambda_A \otimes {\cal I}$
also form a cone $K_{AB}$ that contains both $K_A$ and $K_B$. If $\rho_0$ is a 
faithful separable state, the map (\ref{extendmap}) sends all elements
of $K_{AB}$ into the cone of separable states, and hence detects 
at least the same set of 
separable states than the map $g$ alone. And finally, the search for a map
in $K_{AB}$ that satisfies (\ref{extendmap}) for some state $\sigma$
is indeed a semidefinite program.

\section{Characterizing the set of detected states}

The image of the cone $K_B$ under the map $g$ determines the set
of separable states that can be detected for a particular base state
$\rho_0$. We have seen that the best choice for $\rho_0$ is a faithful
separable state, since in that case the mapping $g$ is one to one
and hence the image of $K_B$ is as large as possible. This leads us to the
obvious question of how big this cone is, or how much of the set of separable
states can be detected in this way. So far we know that the
set of detectable separable states (for a fixed faithful base state) is a cone
of nonzero measure contained in the cone of separable states. We do not 
expect it to be the whole cone of separable states, since the separability
problem is NP-hard and our characterization is polynomial.  
However we will show that we can detect in this way any state in the
interior of the set of separable states $S$, 
except maybe for a set of measure zero. 

One set of states that we can assure will be detected is the
set of separable faithful states in the interior of $S$, which is a set
whose complement (with respect to $S^\circ$) has zero measure. 
Let $K^+(\rho_0) = \{ \sigma : \sigma = 
{\cal I} \otimes \Lambda (\rho_0), \mathrm{with}\ \Lambda 
\ \mathrm{a \ positive\  map}\}.$
This is also a cone and it is contained in the cone
of states characterized by the SDP discussed in the previous
section. Even though this is a smaller set of states, it is
enough to prove all the properties we want to present. Note that it
is clear that for any faithful state $\rho_0$ (actually any state),
$\rho_0 \in K^+(\rho_0)$, since we can just take $\Lambda$ to be
the identity map. But this observation is not at all helpful
since we need to know if $\rho_0$ is actually separable in order
for this approach to be useful. Fortunately, for faithful states
in $S^\circ$ we can prove a stronger result:
\newtheorem{thm1}{Theorem}
\begin{thm1}
Let $\sigma$ be a faithful state in $S^\circ$. There is another faithful state
$\rho_0 \in S^{\circ}$ such that $\sigma \in (K^+(\rho_0))^\circ$.
\end{thm1}

\noindent \textbf{Proof:} Let $\sigma \in S^{\circ}$. Consider the linear map
$h(\Lambda) = {\cal I} \otimes \Lambda (\sigma)$. By the continuity of
this map, there is a neighborhood $A$ of the identity map ${\cal I}$
such that if $\Lambda \in A \ \Rightarrow \ h(\Lambda) \in S^\circ$.
Furthermore, since the identity map is invertible, there is a nighborhood
$B \subseteq A$ such that all maps in $B$ are also invertible. Let $inv$ be
the inverse function, that sends any invertible map to its inverse. Then
it is clear that $inv = inv^{-1}$. This together with the fact that $inv$ is a 
continuous function tells us that $inv$ send open sets into open sets.
Consider then the open set $B^{-1} = inv(B)$. Since ${\cal I} \in B$ and
${\cal I}^{-1} = {\cal I}$, then ${\cal I} \in B^{-1}$. On the other hand,
the identity map is on the boundary of the set of positive maps since
an arbitrarily small perturbation of this map can send a PSD operator
into a non-PSD operator. Therefore 
the intersection of $B^{-1}$ with the interior of the set of positive maps
is a nonempty open set $C$. Now consider the open set $C^{-1} = inv(C)$. 
Clearly, $C^{-1} \subset A$ and every element $\Lambda$ of $C^{-1}$ is a map
that satisfies \textit{(i)} $\Lambda \in A$ and hence 
$h(\Lambda) \in S^\circ$, and
\textit{(ii)} $\Lambda$ is invertible and $\Lambda^{-1}$ is a positive map.
Finally, consider the set $h(C^{-1}) \subset S^\circ$. This is a nonempty
open subset of $S^\circ$, and so it must contain a faithful state
$\rho_0$ (different from $\sigma$ since the identity map is
not in $C^{-1}$), because faithful states are dense in the set of states. Then
there is a map $\bar\Lambda^{-1}$ in $C^{-1}$ such that 
$\rho_0 = {\cal I} \otimes 
\bar\Lambda^{-1} (\sigma)$. Therefore the map $\bar\Lambda \in C$
satisfies $\sigma = {\cal I} \otimes \bar\Lambda (\rho_0)$, with
$\bar\Lambda$ positive. Since $C$ is an open set contained
in the interior of the set of positive maps, and the
function $g(\Lambda) = {\cal I} \otimes \Lambda (\rho_0)$ is linear
and one to one,  $\sigma$ must belong to $(K^+(\rho_0))^\circ$. $\Box$

The preceding theorem tells us that any separable faithful state
can be detected using a different separable faithful state as
a base state. This comprises all of the set of separable states
except for a set of measure zero (namely, the set of nonfaithful states
and the states in the boundary of $S$ that we have left out of this
discussion.) We will see that some nonfaithful states are also detected
by this procedure. To understand this we need the following lemma:
\newtheorem{lemma}{Lemma}
\begin{lemma}
Let $\rho_0$ be a faithful state in $S$. The state
$\sigma = {\cal I} \otimes \bar\Lambda (\rho_0)$ with $\bar\Lambda$ 
a positive map
is also separable and faithful if and only if the map $\bar\Lambda$ 
is invertible.
\end{lemma}

\noindent \textbf{Proof:} Consider the function $f(\Lambda) =
{\cal I} \otimes \Lambda (\sigma)$. The state $\sigma$ is faithful
if and only if this function is one to one. We can rewrite $f$ as
$f(\Lambda) = {\cal I} \otimes (\Lambda \circ \bar\Lambda) (\rho_0)$.
If $\bar\Lambda$ is invertible then  $f$ is one to one, since 
$f(\Lambda_1) = f(\Lambda_2)$ implies that $\Lambda_1 \circ \bar\Lambda
= \Lambda_2 \circ \bar\Lambda$ because $\rho_0$ is faithful, 
and so $\Lambda_1 = \Lambda_2$ because $\bar\Lambda$ is invertible.
To prove the converse, assume that $\sigma$ is faithful and that
$\bar\Lambda$ is not invertible, so that $\mathrm{Ker}(\bar\Lambda) 
\neq \emptyset$.
We can choose two maps $\Lambda_1$ and $\Lambda_2$, $\Lambda_1 \neq
\Lambda_2$ such that
they are distinct only over $\mathrm{Ker}(\bar\Lambda)$. Then we have that
$\Lambda_1 \circ \bar\Lambda = \Lambda_2 \circ \bar\Lambda$.
Now we can write 
\bea
f(\Lambda_1) &=& {\cal I} \otimes \Lambda_1 (\sigma) \nonumber \\
             &=& {\cal I} \otimes (\Lambda_1 \circ \bar\Lambda) (\rho_0) 
\nonumber \\ 
             &=& {\cal I} \otimes (\Lambda_2 \circ \bar\Lambda) (\rho_0) 
\nonumber \\  
             &=& {\cal I} \otimes \Lambda_2 (\sigma) = f(\Lambda_2),
\eea
which is a contradiction to the fact that $f$ is one to one since
$\sigma$ is faithful. Thus, $\bar\Lambda$ must be invertible. $\Box$

This result shows that there are nonfaithful states that are detected,
since we can easily construct a noninvertible positive map $\Lambda$ and
hence the state ${\cal I} \otimes \Lambda (\rho_0)$ is not faithful
and belongs to $K^+(\rho_0)$ for $\rho_0$ faithful. However 
this is not enough to say that all nonfaithful states can be detected.
Whether there are nonfaithful states that cannot be detected in this way is an
interesting and open problem at this point. 

Another interesting result is given by the following lemma:
\begin{lemma}
Let $\rho_1$ and $\rho_2$ be faithful states such that
$\rho_2 \in K^+(\rho_1)$. Then $K^+(\rho_2) \subseteq K^+(\rho_1)$.
\end{lemma}
 
\noindent \textbf{Proof:} Let $\sigma \in K^+(\rho_2)$. Then
$\sigma = {\cal I} \otimes \Lambda (\rho_2)$ for some positive map
$\Lambda$. Since $\rho_2 \in K^+(\rho_1)$, there
is a positive map $\Lambda'$ such that 
$\rho_2 = {\cal I} \otimes \Lambda' (\rho_1)$. Combining the two equations we
have that $\sigma = {\cal I} \otimes (\Lambda \circ \Lambda') (\rho_1)$.
Since the composition of two positive maps is also a positive map,
then $\sigma \in K^+(\rho_1)$, and hence $K^+(\rho_2) \subseteq K^+(\rho_1)$.
$\Box$

An interesting picture of the set of separable states begins to emerge.
By combining Theorem 1 and Lemma 2 we can conclude that for any 
faithful state $\rho_2$ in $S^\circ$ we can always find another faithful
state $\rho_1$ also in $S^\circ$ such that $K^+(\rho_2) \subseteq K^+(\rho_1)$.
We could then construct a sequence of cones $K^+(\rho_n)$ such that
$K^+(\rho_{n}) \subseteq K^+(\rho_{n+1})$. We do not expect this sequence
of cones to contain all possible detectable states asymptotically, but
rather to grow to some sort of maximal cone that will define
a domain inside the set of separable states. Since this domain cannot
be the whole set $S$, we expect that there will be other domains
constructed in the same way, using faithful base states that are not
contained in any of the cones $K^+(\rho_n)$. These domains must cover
the whole interior of $S$ except maybe for a set of measure zero,
comprised only of nonfaithful states. 

To summarize, the procedure for detecting separable states could be applied
as follows.
Given a state $\sigma$ we want to check for separability, we first
generate a random separable state by randomly generating the
terms in the decomposition 
\be
\rho_0 = \sum_i p_i \, \rho_A^{(i)} \otimes \rho_B^{(i)},
\ee
i.e., randomly choosing the number of terms in the decomposition,
the probabilities $p_i$ and the local density matrices 
$\rho_A^{(i)}$ and $\rho_B^{(i)}$. We can simplify this step by noting
any separable state can be written as a convex combination
of pure product states with at most $d^4+1$ terms. Then we check
whether this random separable state is faithful or not 
by checking if the operator $\check{\rho}_0 = (E \rho_0)^{T_B} E$ 
is invertible  (with $E=\sum_{ij} 
|ij\rangle\langle ji|$ the swap operator.) Since the set of nonfaithful states
has zero measure, our random state will be most likely
faithful. Then, we can just solve the enhanced SDP 
discussed in Section \ref{enhSDP}, and if a feasible
solution exists, we have proven that $\sigma$ is separable, and can
furthermore construct an explicit convex decomposition of it in terms
of product states. If the SDP turns out to be unfeasible, 
it means that $\sigma$ could be entangled, or that we need a
different faithful base state. We could then select a 
new faithful base state and proceed as before.

Since each faithful state can help us to identify many separable
states, it could be useful to save them for use with other
states. We can then build up a table of faithful separable states
for each dimensionality, so we can bypass the random generation
step in the future. Furthermore, this table can be
optimized by discarding any state $\rho_2$ that belongs
to $K^+(\rho_1)$ for some $\rho_1$ also in the table,
since according to Lemma 2 it will not detect any new states. 
Another interesting and open problem is whether we can make this
table finite and still detect most of the states in $S^\circ$ except maybe
for a set of arbitrarily small (but not zero) measure. That finite
number of base states will probably depend on the dimensions of the problem
(and the measure of the undetected set), but this approach might turn out to be
a useful tool for low dimensional problems.

\subsection{Extension to $d_A \neq d_B$}

Our approach to detect separable states can be extended 
to the case in which the dimensions 
of the spaces ${\cal H}_A$ and ${\cal H}_B$ are not equal, with 
a few minor modifications. To see how this works we first need
to state the Jamio{\l}kowski isomorphism in its more general form.
Let  ${\cal H}_A$ and ${\cal H}_B$ be two spaces of dimensions
$d_A$ and $d_B$ respectively. We will denote by ${\cal B}({\cal H})$ the
set of bounded, PSD operators over the Hilbert space ${\cal H}$. 
Let ${\cal L}({\cal B}({\cal H}_A),{\cal B}({\cal H_B}))$ be the set
of CP maps from ${\cal B}({\cal H}_A)$ to ${\cal B}({\cal H_B})$. Then
the function $F:{\cal L}({\cal B}({\cal H}_A),{\cal B}({\cal H}_B))
\rightarrow {\cal B}({\cal H}_A \otimes {\cal H}_B)$, defined by
\be
F(\Lambda) = {\cal I} \otimes \Lambda (|\psi^+\rangle_{AA}\,{}_{AA}\langle
\psi^+ |),
\ee
where $|\psi^+\rangle_{AA}$ is the maximally entangled state in 
${\cal H}_A \otimes {\cal H}_{A}$, is an \emph{isomorphism} between
CP maps in ${\cal L}({\cal B}({\cal H}_A),{\cal B}({\cal H}_B))$
and PSD operatos in ${\cal H}_A \otimes {\cal H}_B$. 

Now let us define another  function $\tilde{F}:{\cal L}({\cal B}({\cal H}_A),
{\cal B}({\cal H}_B))
\rightarrow {\cal B}({\cal H}_A \otimes {\cal H}_B)$, given by
\be
\tilde{F}(\Lambda) = {\cal I} \otimes \Lambda (\rho_0),
\ee    
with $\rho_0$ a faithful separable state in ${\cal H}_A \otimes {\cal H}_{A}$.
It is clear that $\tilde{F}(\Lambda)$ is also a separable state
in ${\cal H}_A \otimes {\cal H}_{B}$ provided that $\Lambda$ satisfies
the same positivity conditions we required in the case $d_A = d_B$.
By the same isomorphism, applied to states in 
${\cal H}_A \otimes {\cal H}_{A}$, we know that there must be a CP map 
$\Lambda_0 : {\cal B}({\cal H}_A) \rightarrow {\cal B}({\cal H}_A)$ such that
\be
\rho_0 = {\cal I} \otimes \Lambda_0 (|\psi^+\rangle_{AA}\,{}_{AA}\langle
\psi^+ |).
\ee
Furthermore, following the same reasoning as in the proof of Lemma 1, 
we conclude that the map $\Lambda_0$ must be invertible (note that
this inverse will not be in general a CP map.) And using 
the same argument once again, it is not difficult to show that
the function $\tilde{F}$ must be one to one. By the same arguments
of the case $d_A = d_B$, we can then conclude that the set of
detected states is also of nonzero measure when $d_A \neq d_B$.
The only difference between these two cases is that in the latter
the faithful states must be chosen from a space different than
the one in which the states to be detected live.   

\subsection{Multipartite states}

It is interesting to see what happens with this approach when we
apply it to multipartite states. Even though the basic property
that allows us to connect two separable states is still valid,
the results are not as useful because the set of detected
states has zero measure even for faithful states. Let us briefly
show the reason in the tripartite case. Let $\rho_0$ be a 
separables state in ${\cal H}_A \otimes {\cal H}_A \otimes {\cal H}_A$,
with $ \mathrm{dim}\, {\cal H}_A = d$.
Clearly, any state of the form $\sigma = [{\cal I} \otimes
{\cal I} \otimes \Lambda] (\rho_0)$, is separable if $\Lambda$ satisfies
the required positivity constraints. The cone of separable tripartite
states is embedded in a vector space of dimension $d^8$. But the
set of linear maps $\Lambda$ is embedded in a space of dimension $d^4$, so
even if $\rho_0$ is a faithful state, the mapping of $\Lambda$ to $\sigma$
produces a set of dimension $d^4$ embedded in a space of dimension $d^8$.
Hence, the image of this mapping (which is the set of detected states) 
has zero measure with respect to the cone of tripartite separable
states, and this makes this approach not useful in practice. One solution
to this problem will be to consider transformations of the form
$\sigma = [{\cal I} \otimes  \Lambda_A \otimes {\Lambda_B}](\rho_0)$
but then checking for the existence of the maps $\Lambda_A$ and $\Lambda_B$
becomes a nonlinear problem that cannot be solved using SDP.

\section{Analytical criterion}

This technique can also help us to derive some analytical results
about separability. By choosing a particular base state, and using some results
on the characterization of positive maps~\cite{doherty2003d}, we can
obtain a sufficient condition for separability in terms of 
the eigenvalues of an associated state.

Consider the following family of states 
\be
\rho(\lambda) = (1-\lambda) I_{AB} + \lambda |\psi^+\rangle \langle
\psi^+ |,
\ee
where $0\le \lambda \le 1$, $I_{AB}$ is the maximally mixed state in
${\cal H}_A \otimes {\cal H}_B$ (i.e., the identity matrix times
$\frac{1}{d^2}$, since we are considering $\mathrm{dim}\, {\cal H}_A = 
\mathrm{dim}\, {\cal H}_B = d$), and $|\psi^+\rangle = \frac{1}{\sqrt{d}}
\sum_i |ii\rangle_{AB}$ is the maximally entangled state. These states
are referred to as the \emph{isotropic states} (since they are
invariant under $U\otimes U^\ast$ transformations) and are known
to be separable for $0\le \lambda \le \frac{1}{d+1}$, and entangled
otherwise. We will choose $\rho_0 = \rho(\frac{1}{d+1})$.
This state is in the boundary of $S$ and it is not difficult
to check that it is faithful.

Let $\sigma$ be a state that satisfies
\be
\sigma = {\cal I} \otimes \Lambda (\rho_0),
\ee
for some positive map $\Lambda$. This is equivalent to
\be
\label{sigma}
\sigma = {\cal I} \otimes \Lambda (\frac{d}{d+1} I_{AB}) +
{\cal I} \otimes \Lambda (\frac{1}{d+1} |\psi^+\rangle \langle
\psi^+ |).
\ee
Let us impose a restriction on $\Lambda$ and require $\Lambda (I_B)=
 I_B$. We will see
later that this restriction can be eliminated, but for now it simplifies
the reasoning. Then we can rewrite (\ref{sigma}) as
\be
\label{sigma2}
\sigma = \frac{d}{d+1} I_{AB} +
\frac{1}{d+1} {\cal I} \otimes \Lambda (|\psi^+\rangle \langle
\psi^+ |).
\ee
The restriction on $\Lambda$ means that, for consistency, we must have
$\mathrm{Tr}_A [\sigma] = I_B$. Again, this restriction can be 
lifted and we will see that later on. We now write
\be
\label{sig-Id}
\sigma - \frac{d}{d+1} I_{AB} =
{\cal I} \otimes \tilde{\Lambda} (|\psi^+\rangle \langle
\psi^+ |),
\ee
where we have rescaled the positive map, so $\tilde{\Lambda} = \frac{1}{d+1}
\Lambda$. 

From the Jamio{\l}kowski isomorphism we also know that any operator $Z$ 
of the form
\be
\label{Z}
Z = {\cal I} \otimes \tilde{\Lambda} (|\psi^+\rangle \langle
\psi^+ |),
\ee
with $\tilde{\Lambda}$ a positive map, must be positive over the set of
separable states. This is equivalent to the associated bihermitian
form 
\bea
E_Z & = & \langle x y|Z|x y\rangle \nonumber \\
    & = & \sum_{ijkl} x_i^\ast y_j^\ast x_k y_l \langle ij|Z|kl\rangle,
\eea
being nonnegative. In~\cite{doherty2003d}
a sequence of tests was introduced (each test implementable as a SDP) that
could prove the positivity of the form $E_Z$, and hence the 
existence of a positive map satisfying (\ref{Z}). 

Consider our state $\sigma$ that
satifies $\mathrm{Tr}_A [\sigma] = I_B$ and let us assume that we can prove
that the bihermitian form associated with the operator
$(\sigma -\frac{d}{d+1} I_{AB})$ is positive, using the techniques 
from~\cite{doherty2003d}. Then,
by the Jamio{\l}kowski isomorphism, the map $\tilde{\Lambda}$ that satisfies
\be
\sigma - \frac{d}{d+1} I_{AB} =
{\cal I} \otimes \tilde{\Lambda} (|\psi^+\rangle \langle
\psi^+ |),
\ee   
is positive. By tracing over $A$ on both sides it is not difficult to
show that $\tilde{\Lambda} (I_B) = \frac{1}{d+1} I_B$, so we can write
\be
\sigma - \frac{d}{d+1} I_{AB} =
\frac{1}{d+1}{\cal I} \otimes \Lambda (|\psi^+\rangle \langle
\psi^+ |),
\ee
where now $\Lambda$ satisfies $\Lambda (I_B) = I_B$. Since $I_{AB} = 
I_A \otimes I_B$ we can write $I_{AB} = {\cal I} \otimes \Lambda (I_{AB})$
and then we have that
\bea
\sigma & = & {\cal I} \otimes \Lambda (\frac{d}{d+1} I_{AB}) +
{\cal I} \otimes \Lambda (\frac{1}{d+1} |\psi^+\rangle \langle
\psi^+ |) \nonumber \\
 & = & {\cal I} \otimes \Lambda (\rho_0),
\eea
which shows that $\sigma$ is separable. 

Assume now that we have a state $\sigma$ that satisfies  
$\mathrm{Tr}_A [\sigma] = \sigma_B > 0$, so that $\sigma_B$ is invertible.
Note that any state in the interior of $S$ will have this
property. We can construct the state
\be
\label{sigmat}
\tilde{\sigma} = \frac{1}{d} (\mathbf{1}_A \otimes \sigma_B^{-\half})\,
\sigma \,
(\mathbf{1}_A \otimes \sigma_B^{-\half}),
\ee
that satisfies the condition $\mathrm{Tr}_A [\tilde{\sigma}] = I_B$
(with $\mathbf{1}_A$ the identity matrix in $d$ dimensions).
It is clear that if $\tilde{\sigma}$ is separable, so is $\sigma$.

According to our discussion above, we can prove separability of 
$\tilde{\sigma}$ by proving positivity of the bihermitian
form associated with the operator $\tilde{\sigma} - \frac{d}{d+1} I_{AB}$.
Clearly, a sufficient (but not necessary) condition for this to be true
is the that the operator $\tilde{\sigma} - \frac{d}{d+1} I_{AB}$ itself
be positive semidefinite. If $\lambda_i$ are the eigenvalues
of $\tilde{\sigma}$, this is equivalent to requiring that
$\mathrm{min}_i \,\lambda_i \geq \frac{1}{d(d+1)}$. We then have the 
following corollary:
\newtheorem{corollary}{Corollary}
\begin{corollary}
Let $\sigma$ be a state in ${\cal H}_A \otimes {\cal H}_B$ with
$\mathrm{dim}\, {\cal H}_A = 
\mathrm{dim}\, {\cal H}_B = d$, that satisfies 
$\mathrm{Tr}_A [\sigma] = \sigma_B > 0$.
Let $\tilde{\sigma} = \frac{1}{d} (\mathbf{1}_A \otimes 
\sigma_B^{-\half})\,\sigma \,
(\mathbf{1}_A \otimes \sigma_B^{-\half})$, with eigenvalues $\lambda_i$.
If
\be
\mathrm{min}_i \,\lambda_i \geq \frac{1}{d(d+1)},
\ee
then $\sigma$ is separable.
\end{corollary} 

Furthermore, for any state $\sigma$ shown to be separable 
as a consequence of Corollary 1, we can provide an explicit 
convex decomposition in terms of product states. 
From the Jamio{\l}kowski isomorphism we can also extract an explicit
expression of the map $\Lambda$ in terms of the hermitian operator
$Z$ that satisfies $Z={\cal I} \otimes \Lambda 
(|\psi^+\rangle \langle\psi^+ |)$. In our case, given an orthonormal product 
basis $\{ |ij\rangle_{AB} \}$,
we can write
\be
\label{lambdarho}
\Lambda (\rho) = \sum_{ijkl} \langle ij|(\tilde{\sigma} - \frac{d}{d+1} I_{AB})
|kl\rangle \langle i|\rho|k\rangle
|j\rangle \langle l|,
\ee
where $\tilde{\sigma}$ is given by (\ref{sigmat}).
Now we just need a decomposition of our base state 
\be
\rho_0 = \frac{d}{d+1} I_{AB} + \frac{1}{d+1} |\psi^+\rangle \langle
\psi^+ |.
\ee
In~\cite{rungta2002a} one such decomposition of $\rho_0$ was
given and we will follow that construction. 
First, let us define a vector ${\mathbf z} = (z_1,\dots,z_d)$, whose
components $z_j$ take on the values $\pm 1$ and $\pm i$.
To each vector ${\mathbf z}$ we associate a pure state 
\be
|\Phi_{\mathbf z} \rangle = \frac{1}{\sqrt{d}} \sum_{j=1}^d z_j |j\rangle,
\ee
with $\{ |j\rangle \}$ the canonical basis. There are $4^d$ such vectors,
and hence $4^d$ states $|\Phi_{\mathbf z} \rangle$, although only
$4^{d-1}$ are distinct in that they differ by more than a global phase.
Now we define a product state in ${\cal H}_A \otimes {\cal H}_B$ given
by
\be
\rho_{\mathbf z} = |\Phi_{\mathbf z} \rangle \langle \Phi_{\mathbf z}|
\otimes |\Phi_{{\mathbf z}^\ast} \rangle \langle \Phi_{{\mathbf z}^\ast}|.
\ee
The ensemble consisting of all $4^d$ of these states with the same
probability gives the density operator
\be
\frac{1}{4^d} \sum_{{\mathbf z}} \rho_{\mathbf z} =
\frac{1}{4^d d^2} \sum_{jklm} \(\sum_{{\mathbf z}} z_j z_k^\ast
z_l^\ast z_m \) |j\rangle \langle k| \otimes |l\rangle \langle m|.
\ee
Since
\be
\sum_{{\mathbf z}} z_j z_k^\ast z_l^\ast z_m = 4^d (\delta_{jk}
\delta_{lm} + \delta_{jl} \delta_{km} - \delta_{jk} \delta_{lm} 
\delta_{jl}),
\ee
it follows that 
\be
\frac{1}{4^d} \sum_{{\mathbf z}} \rho_{\mathbf z} = 
I_{AB} + \frac{1}{d} |\psi^+\rangle \langle
\psi^+ | - \frac{1}{d^2} \sum_{j=1}^d |j\rangle \langle j| \otimes
|j\rangle \langle j|.
\ee
Multiplying by $\frac{d}{d+1}$ and rearranging terms we get
\be
\label{rho0dec}
\rho_0 = \frac{d}{d+1} \frac{1}{4^d} \sum_{{\mathbf z}} \rho_{\mathbf z}
+ \frac{1}{(d+1)d} \sum_{j=1}^d |j\rangle \langle j| \otimes
|j\rangle \langle j|,
\ee
and hence we have 
\bea
\label{sigmadec}
\sigma &=&  \frac{d^2}{d+1} \frac{1}{4^d} \sum_{{\mathbf z}} 
|\Phi_{\mathbf z} \rangle \langle \Phi_{\mathbf z}|
\otimes \sigma_B^{\half}\Lambda(|\Phi_{{\mathbf z}^\ast} \rangle 
\langle \Phi_{{\mathbf z}^\ast}|)\sigma_B^{\half} + \nonumber \\
& & + \frac{1}{(d+1)} \sum_{j=1}^d |j\rangle \langle j| \otimes
\sigma_B^{\half}\Lambda(|j\rangle \langle j|)\sigma_B^{\half},
\eea
where $\sigma_B = \mathrm{Tr}_A [\sigma]$ and $\Lambda$ is
given by (\ref{lambdarho}). Since the map $\Lambda$ is positive,
Eq. (\ref{sigmadec}) gives an explicit convex decomposition of $\sigma$ in
terms of product states, certifying its separability.

It is interesting to compare this corollary to the well-known
result of Gurvits and Barnum~\cite{gurvits2002b} that characterizes
the biggest ball of separable states centered on the maximally mixed state.
Their result says that if $\parallel \sigma - I_{AB} 
\parallel_2^2 \leq \frac{1}{d^2 (d^2 -1)}$, then the state $\sigma$ is
separable. Now consider a state that satisfies 
$\mathrm{Tr}_A [\sigma (\epsilon)] = I_B$ in the basis in which it is
diagonal, and that
\be
\label{sigdiag}
\sigma (\epsilon) = \mathrm{diag}( \epsilon + \frac{1}{d(d+1)}, \lambda, 
\ldots, \lambda),
\ee
where $\lambda = \frac{1}{d(d+1)} + \delta$ and both $\epsilon$ and
$\delta$ are positive. By normalization we must have $\frac{d^2}{d(d+1)}
+\epsilon + (d^2-1) \delta =1$, and hence
\be
\delta = (1-\epsilon - \frac{d}{d+1})\frac{1}{d^2-1}.
\ee
Since we need $\delta \ge 0$, we need $\epsilon \le 1-\frac{d}{d+1}$.
If these conditions are satisfied, the state $\sigma (\epsilon)$ is separable.
To see what the Gurvits-Barnum criterion says about such a state,
we define the function
\be
\label{f}
f(d,\epsilon) = \parallel \sigma (\epsilon) - I_{AB} 
\parallel_2^2 -\frac{1}{d^2 (d^2 -1)}.
\ee
If $f(d,\epsilon) > 0$ for $0 < \epsilon \le 1-\frac{d}{d+1}$, then
the state $\sigma (\epsilon)$ is not shown as separable by the Gurvits-Barnum
criterion, but it is by ours. It is easy to check that
\be
f(d, 1-\frac{d}{d+1}) = \frac{d-2}{d^3 - d} > 0, \ \ \ \forall d \ge 3.
\ee
Since $f(d,\epsilon)$ is a continuous function of $\epsilon$, this shows that 
for $d \ge 3$, there is always a range of values of $\epsilon$ for
which the state $\sigma (\epsilon)$ is shown to be separable by our technique
but not by the Gurvits-Barnum criterion.

\section{Conclusions}

In this paper we have introduced a new technique to
detect separable states. The idea is to show
that two states, the first of which is known to be separable
and has certain properties,
are connected by a local map that preserves separability. 
This local map is required to be positive on a set of
local states determined by the convex decomposition
of the first state in terms of product states. 
If a map with the required properties is found connecting the
two states, the second state is proven to be separable.
The key point is that searching for this
connecting map reduces to a semidefinite program (SDP)
which can be implemented efficiently. Furthermore, if the
second state is proven to be separable, the solution of the
SDP provides the required map and an explicit proof of
separability of the second state can be constructed in the form 
of a convex decomposition in terms of product states.

It is important to note that the local map in question
needs not be positive over all states. This is an advantage,
since characterizing the set of positive maps is an NP-hard
problem, while the wider set of maps we are interested in (which includes
the set of positive maps) is easier to characterize and 
allows us to reduce the problem to a SDP. However, restricting
ourselves to the set of positive maps makes it easier to prove
and understand some properties of this technique. In particular,
coupled with some previous work aimed at characterizing 
positive maps via an infinite hierarchy of conditions~\cite{doherty2003d}, it
allowed us to extract a new criterion for separability based
on the spectral properties of an associated state.

We have shown that all states in the interior of the set of separable
states can be detected by this technique, except maybe for a set of 
measure zero. Our approach proceeds by choosing a random separable
state (with a certain property known as 
\textit{faithfulnes}~\cite{d'ariano2003a}) that we call 
\textit{base states},
and implementing a SDP to check whether a given state is separable
or not. The set of states detected using a particular base state
is a convex subset of the set of separable states. By choosing
different base states we are able to detect more and more 
separable states. We could keep a table of these states 
for future use much in the same way we keep the information
about entanglement witnesses. Even though an infinite number
of these states are needed to characterize all separable states,
it may well be the case that a finite number will be enough
to characterize almost the whole set, except maybe for a set of
very small (but not zero) measure. This is still an open question. 
Also open for further research is the structure of
the set of separable states that cannot be detected 
using this technique. 
This structure seems to be related to the
structure of the set of non invertible maps.  

Finally, it is worth stressing that, as many other
separability criteria, our technique is not complete, and
should be complemented with other ways of analyzing entanglement
and separability. One feature that should be pointed out is that
our criteria either proves separability of the state (and
provides a certificate of it in the form of an explicit decomposition) or
fails. Other criteria usually show that the state is \textit{entangled}
or fail. These criteria generate an approximate characterization
of the set of separable states from the outside. Our approach
works \textit{from the inside}. In this respect it is similar to the
algorithm presented in~\cite{hulpke2005a}, that generates a sequence of 
convex sets included in the set of separable states for which membership
can be easily checked.  These two types of criteria (inside and outside 
characterizations)
should be used together to increase the chance of correctly
identifying the entanglement properties of a given state.

\section{Acknowledgements}

I would like to thank Vwani Roychowdhury for his support and suggestions,
and Pablo Parrilo for many useful discussions. This work was supported by
the MARCO/FCRP program, and the Western Institute of Nanoelectronics, a 
Nanoelectronics Research Initiative.

\bibliographystyle{prsty}
\bibliography{Cav}

\end{document}